\documentclass[twocolumn,pre,showpacs,preprintnumbers,amsmath,amssymb,nofootinbib,superscriptaddress,floatfix]{revtex4-1}
\pdfoutput=1
\usepackage{amsfonts}
\usepackage{amsmath}
\usepackage{amssymb}
\usepackage{graphicx}
\usepackage{subcaption}
\usepackage{dcolumn}
\usepackage{bm}
\usepackage[usenames, dvipsnames, table]{xcolor}
\newcommand{\lp}{\left (}
\newcommand{\rp}{\right )}

\newcommand{\ra}{\rightarrow}

\newcommand{\ev}[1]{\left \langle #1 \right \rangle}

\newcommand{\beq}{\begin{equation}}
\newcommand{\eeq}{\end{equation}}

\newcommand\mnras{MNRAS}%
\begin{document}
% Use the \preprint command to place your local institutional report
% number in the upper righthand corner of the title page in preprint mode.
% Multiple \preprint commands are allowed.
% Use the 'preprintnumbers' class option to override journal defaults
% to display numbers if necessary
%\preprint{}
%Title of paper
\title{Scaling Relations of Halo Cores for Self-Interacting Dark Matter}
% repeat the \author .. \affiliation etc. as needed
% \email, \thanks, \homepage, \altaffiliation all apply to the current
% author. Explanatory text should go in the []'s, actual e-mail
% address or url should go in the {}'s for \email and \homepage.
% Please use the appropriate macro foreach each type of information
% \affiliation command applies to all authors since the last
% \affiliation command. The \affiliation command should follow the
% other information
% \affiliation can be followed by \email, \homepage, \thanks as well.
%\author{}
%\email[]{Your e-mail address}
%\homepage[]{Your web page}
%\thanks{}
%\altaffiliation{}
%\affiliation{}
\author{Henry W. Lin}
%\email[]{}
\affiliation{Institute for Theory \& Computation, Harvard-Smithsonian Center for Astrophysics, 60 Garden Street, Cambridge, MA 02138, USA}
\author{Abraham Loeb}
\affiliation{Institute for Theory \& Computation, Harvard-Smithsonian Center for Astrophysics, 60 Garden Street, Cambridge, MA 02138, USA}
\date{\today}
\begin{abstract}
Using a simple analytic formalism, we demonstrate that significant dark matter self-interactions produce halo cores that obey scaling relations nearly independent of the underlying particle physics parameters such as the annihilation cross section and the mass of the dark matter particle. For dwarf galaxies, we predict that the core density $\rho_c$ and the core radius $r_c$ should obey $\rho_c r_c \approx 41 \,\text{M}_\odot \text{pc}^{-2}$ with a weak mass dependence $\sim M^{0.2}$. Remarkably, such a scaling relation has recently been empirically inferred. Scaling relations involving core mass, core radius, and core velocity dispersion are predicted and agree well with observational data. By calibrating against numerical simulations, we predict the scatter in these relations and find them to be in excellent agreement with existing data. Future observations can test our predictions for different halo masses and redshifts.
\end{abstract}
\maketitle
% body of paper here - Use proper section commands
% References should be done using the \cite, \ref, and \label commands
	\paragraph*{Introduction.} The standard cosmological model with a cosmological constant $\Lambda$ and cold dark matter (CDM) has been confirmed on large scales by a wealth of successful predictions. Yet despite its simplicity and experimental success, a variety of challenges to the $\Lambda$CDM paradigm persist. On the observational side, the CDM paradigm suffers from the well-known cusp-core problem (see \cite{deb10} for a review); numerical simulations \cite{nfw97, moore99} predict that the density of dark matter halos should rise sharply towards the center of the halo forming a ``cusp'', whereas observations (see e.g. \cite{spano08, donato09, korm, burkert15} and references therein) indicate that the dark matter in the center of halos rises more gently, resembling a ``core'' even in dwarf galaxies where the baryonic content is negligible \cite{walkerloeb}. Motivated by observations, various models of interacting dark matter have been proposed to alleviate this tension including scattering \cite{spergel00, loeb11} and annihilation \cite{kap00}, although some authors, e.g \cite{strig14}, argue that dark matter cores do not appear upon a more careful analysis. Many interesting experiments (for recent progress see e.g. \cite{dienes, kap15} and references therein) have been devised to constrain some combination of mass and interaction cross section. Here, we focus on the model-independent observational signatures of annihilation. Remarkably, we find that scaling relations exist that are generic predictions of self-interactions where any dependence on the cross section $\sigma$ and the particle mass cancel. We can thus make predictions about annihilation signature without introducing any additional parameters beyond what is required in a minimal $\Lambda$CDM cosmology. Throughout this Letter, we assume $h \equiv H_0/(100 \text{\,km\,s}^{-1}\,\text{Mpc}^{-1})= 0.7$, $\Omega_m = 0.27$, and $\Omega_\Lambda = 0.73$ \cite{planck15}.
	
	\paragraph*{Core scaling relations} To derive the core scaling relations, we need two simple ingredients. The first ingredient is the initial condition of the dark matter density profile, before enough cosmic time has passed for annihilation to significantly alter the profile. We take the dark matter profiles to be of the Navarro-Frenk-White (NFW) form \cite{nfw97}:
	\beq
	\rho(r,t=t_0) = \frac{\rho_0}{(r/r_s)(1+r/r_s)^2}.
	\eeq
	Here $t_0$ is the age of the universe when the dark matter halo virialized. We will comment shortly on the modification of our results if the inner profile of the dark matter halo is more accurately described by the Einasto profile \cite{einasto}, which is likely a better fit to simulations \cite{merr06}. To determine $\rho_0$ and $r_s$ for a halo of mass $M$ and redshift $z$, we adopt an expression for the concentration parameter $c_{200}(M,z) \equiv r_s/r_{200}$, where $r_{200}$ is defined as the radius interior to which the average density is 200 times the critical density of the universe $\rho_\text{crit}= (3 H^2/8\pi G)$, where $H$ is the Hubble parameter. 

	We adopt the fitting formulae of \citet{dutton14}, their equations 7, 10, and 11, calibrated from the results of their N-body simulations:
	\beq
	\begin{split}
	\log c_\text{200} = a + b \log (M/10^{12} h^{-1} M_\odot)\\
	a=0.520 + 0.385 \exp{(-0.617 z^{1.21})}\\
	b = -0.101 + 0.026 z.
	\end{split}
	\eeq
	These formulae are in agreement with a recently proposed universal model of \cite{diemer15}.
	
	The second ingredient is the evolution of the dark matter profiles with time. We will only consider the simplest case of s-wave annihilation where $\ev{\sigma v} = \text{const}$. This is a reasonable approximation for most annihilation processes unless a quantum selection rule prevents s-wave annihilation or the cross section acquires a sharp velocity dependence due to Sommerfeld enhancement. Ignoring the gravitational back reaction, the time derivative of the density is given by
	\beq
	\dot{\rho}(x,t) = -\Gamma \rho^2(x,t),
	\eeq
	where $\Gamma \rho$ has units of inverse time. The solution to this equation is 
	\beq
	\rho(r,t) = \frac{\rho(r,t_0)}{1+f \, \Gamma\, t \rho(r,t_0)},
	\eeq
	where we have inserted a fudge factor $f > 1$ (effectively rescaling $\Gamma$) in anticipation that gravitational back-reaction should change the profile. (If gravitational back-reaction is negligible, $f=1$). 
		
	Plugging in our initial condition yields%For small radii $r \ll r_s$, $\rho(r,t_0) \approx \rho_0 \lp r/r_s\rp^{-1}$ and we obtain
	\beq
	\label{exact}
	\rho(r,t) = \frac{\rho_c}{(r/r_c)(1+r/r_s)^2 +1},
	\eeq
	where we define a core density and core radius $\rho_c \equiv \lp f\, \Gamma\, t\rp^{-1}$ and $r_c \equiv \lp f \,\Gamma \,t \, \rho_0\rp r_s$, respectively. For the moment, we simply define $r_c$ the core radius, we will show shortly that it corresponds approximately to the notion of core radius defined by other authors. 
	
	Since different authors define these quantities differently, it is worth noting what these quantities represent, so that they can be sensibly compared to observational work. It is straightforward to verify that both quantities can be expressed in terms of properties of the central density profile: the central density is simply $\rho_c = \rho(r=0)$, and Taylor expanding the logarithmic slope
\beq
- \frac{d\log \rho}{d\log r} = \lp \frac{r}{r_c}\rp + \mathcal{O}\lp r^2\rp
\eeq
shows that $r_c$ is the length scale over which the logarithmic slope significantly deviates from 0. A full comparison of the behavior of our density profiles with existing work is given in Figure 1, which shows that the numerical value of $r_c$ (obtained by fitting to data) may differ from the values of $(r_c)_\text{Burkert}$ or $(r_c)_\text{Zavala}$, depending on the precise value of $r_s$. However, the difference is unlikely to be greater than $\sim 0.1$ to $0.2$ log increments, which translates to a difference of several tens of percent. In other words, for a large range in possible values of $r_s$, our definition of $r_c$ is consistent with $(r_c)_\text{Burkert}$ or $(r_c)_\text{Zavala}$ up to a factor of order unity.

 These {\it non-parametric} properties of the central density profile allow us to connect our definitions with other parametric profiles. For example, the above considerations lead us to identify $\lp\rho_0\rp_\text{Burkert} = \rho_c$ and $\lp r_c\rp_\text{Burkert} \approx r_c$ where the subscripts denote definitions found in Burkert \cite{burkert15}, since the core radii in both our profile and Burkert's profile are turnover radii, e.g., the length scale over which the logarithmic slope becomes non-negligible.  %For example, if we approximate $d \log \rho/d r = d \log \rho/dr|_{r=0}$, we may derive an approximation for the inner density profile
%\beq
%\rho(r) \approx  \frac{\rho_c}{2} e^{-r/r_c}
%\eeq

Note that we have not made any assumptions about the constancy of $f$ with particle physics or halo parameters. \citet{kap00} have argued that $3 \lesssim f \lesssim 10$ (decreasing with mass) accurately accounts for the gravitational back-reaction in the limit that annihilation occurs slowly such that the adiabatic invariant $r M(r)$ is conserved; numerical simulations of annihilating \cite{kap00,dave} and self-interacting (scattering only) \cite{elbert} dark matter confirm the accuracy of this approximation. These simulations also show that our results are qualitatively unaffected by the choice of NFW or Einasto profiles, at least when the dark matter interaction is strong enough so that the core radius $r_c \gtrsim 10^{-2} \, r_s$ and the differences between NFW and Einasto are significant.

More quantitatively, \citet{zavala} showed that self-interacting dark matter profiles with $\sigma/m = 0.1$ cm$^2$ g$^{-1}$ are well fit by a profile of the form (using our notation)
	\beq
	\rho_\text{Zavala}(r) = \frac{\rho_c}{(r/r_c)(1+r/r_s)^2 +(1+r/r_s)^2}.
	\eeq
For small radii $r \ll r_s$, the last term $(1+r/r_s)^2 \approx 1$ and we recover equation (\ref{exact}). For large radii $r \gg r_s$, the $r^3$ term in the denominator will dominate, and the quadratic term $(1+r/r_s)^2$ can be neglected. In either regime, our model is in agreement with the results of \citet{zavala}. The typical volume averaged difference between our models ${(\rho - \rho_\text{Zavala})}/{\rho_0}$ is not more than a few percent within $2 \, r_s$, and is entirely negligible for much larger volumes.
	%%%% We note that observations \cite{burkert15} of dwarf spheroidal galaxies suggest a core radius $r_c \sim 0.4$ kpc, whereas the NFW scale radius for dwarf galaxies is of order $r_s \sim 3$ kpc. The observationally interesting regime is thus $10^{-2}\, r_s < r_c < r_s$, and hence a reasonable approximation is to neglect $r_c/r_s$ while simultaneously using the NFW profile.
	
	Although we have motivated equation (3) by considering annihilating dark matter, we note that scattering also has a similar effect, as scattering will kick particles from the high density inner regions to larger radii, where their contribution to the mass budget is negligible. Since the scattering rate is also proportional to $\rho^2$, $\Gamma$ in equation (3) should also include contributions from the scattering cross section. These results are in agreement with numerical simulations (e.g. the results of \citet{elbert}) which predict a flat slope interior to a core radius $\sim r_c$ and a profile which converges to the original NFW form at large radii. Note, however, that if a self-interacting dark matter halo is in the gravothermal collapse regime, our results do not apply. Indeed numerical simulations \citep{koch2000, koda11} show that during collapse, the inner density profile becomes significantly more cuspy than a constant density core. The failure of our formalism in this regime is not surprising, as processes such as gravothermal collapse obviously do not preserve adiabatic invariants.
\begin{figure*}
	%\centering % \begin{center}/\end{center} takes some additional vertical space
		\includegraphics[width=0.4\textwidth]{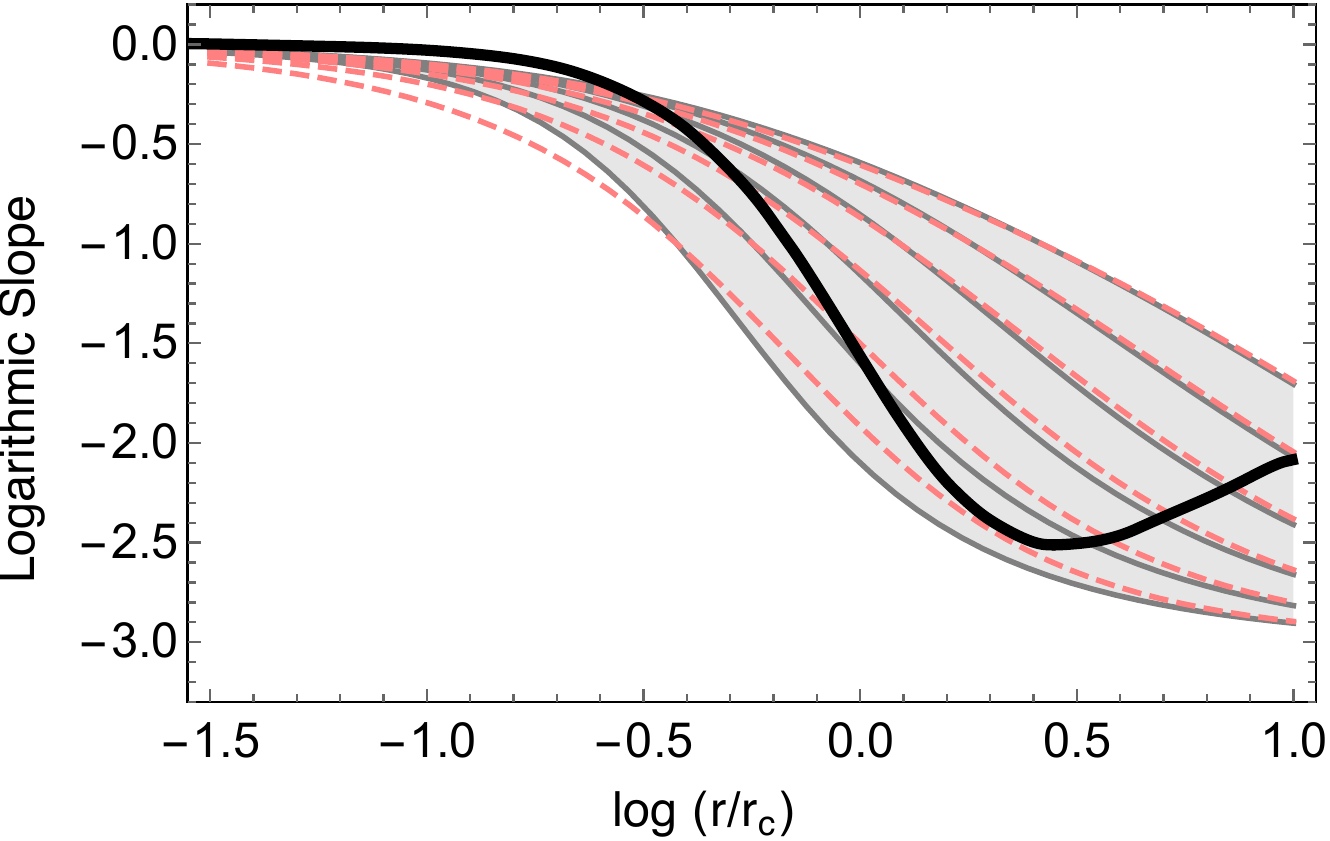}

	\caption{Comparison of our density profile with \citet{burkert15} and \citet{zavala}. On the $y$-axis is the logarithmic slope of the density profile, $d \rho/d \log r$, on the $x$-axis is the logarithmic radius in units of $r_c$. The gray lines show our profile for $r_s = \frac{1}{2} r_c$ to $r_s = 16 \, r_c$ in log increments. Similarly, the pink dashed lines show different values of $r_s$ using the parameterization found in Zavala {\it et. al}. The thick black line is from \cite{burkert15}. Note that observations provide the best constraints in the region $r/r_c \lesssim 1$. This graph illustrates that our definition of $r_c$ is approximately consistent with other definitions of $r_c$ found in the literature. For example, the value of $r_c$ inferred by fitting a density profile to some given data may be somewhat larger or smaller compared to the value of $(r_c)_\text{Burkert}$ or $(r_c)_\text{Zavala}$.}
\end{figure*}

\begin{figure*}
	%\centering % \begin{center}/\end{center} takes some additional vertical space
	\begin{subfigure}[b]{0.4\textwidth}
		\includegraphics[width=\textwidth]{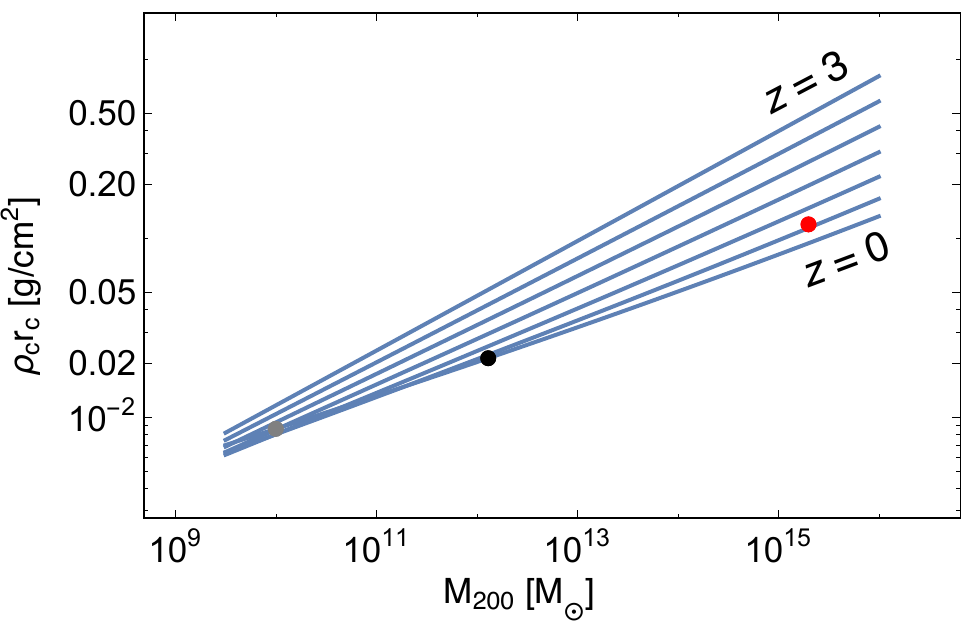}
	\end{subfigure}
	\begin{subfigure}[b]{0.4\textwidth}
		\includegraphics[width=\textwidth]{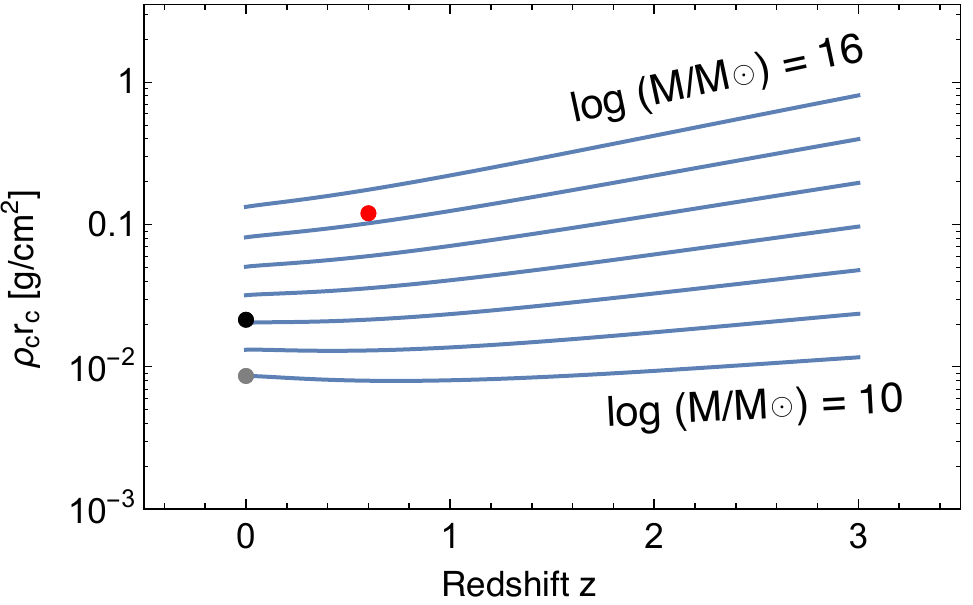}
	\end{subfigure}
	\caption{Predicted $\rho_c r_c$ as a function of mass $M_{200}$ and redshift $z$. As sample points, we display the Milky Way \cite{mcm11} in black, a typical dwarf galaxy in gray, and the Phoenix galaxy cluster \cite{mcdonald} in red. On the right panel, different curves display different halo masses $M_{200}$, which increase in unit log-increments. On the left panel, the different curves represent different redshifts which increase in increments of 0.5. }
\end{figure*}

	Although $r_c$ and $\rho_c$ will have a large spread due to the different formation times of halos, there exists an overall scaling relation
	\beq
	\rho_c r_c = \rho_0 r_s.
	\eeq
	We emphasize that the above equation was not derived by making any assumptions about whether or not $r_c \ll r_s$. Note that the formation time $t$, the fudge factor $f$, and the annihilation rate $\Gamma$ no longer appear in this result for the core surface density. (However if the cross section is negligible, $r_c \ra 0$ and the above scaling relation will be impossible to observe). The quantities on the right hand side of equation (6) are properties of the dark matter halo which can be easily computed. It is a straightforward to integrate equation (1) to find
	\beq
	\rho_0 r_s=\frac{N \, c_{200}^2 \lp \rho_\text{crit} \rp ^{2/3} M^{1/3}}{\log (c_{200}+1)-c_{200}/(1+c_{200}) }.
	\eeq
	where the normalization $N = 10 (10/\pi)^{1/3}/(3^{2/3}) \approx 7.07$. Since $c_\text{200}(M,z)$ is a very weak function of mass, we expect that the right hand side to scale roughly with $M^{1/3}$. A more detailed calculation shows that the mass dependence is  even weaker $\rho_0 r_s \propto M^{0.2}$ at low redshifts and even at a relatively high redshift $z = 3$, $\rho_0 r_s \propto M^{0.3}$.
	Thus, if we consider a subclass of halos such as dwarf galaxies, $\rho_c r_c$ will be approximately constant. 
	This is precisely the scaling relation inferred empirically in \citet{spano08, donato09, korm} and most recently \citet{burkert15}. 
	
	Ultimately, the near-constancy of $\rho_0 r_c$ is a consequence of the fact that the {\it initial} conditions provide by (non-interacting) LCDM simulations have a density profile that scales approximately $\propto r^{-1}$. (The resulting profiles with interaction taken into account do not need $\rho \propto r^{-1}$). For example, if dark matter halos were more accurately described by older secondary infall models \cite{gunn}, the inner profiles would be isothermal: $\rho \propto r^{-2}$, which would not lead to the near-constancy of $\rho_c r_c$.
	
	\paragraph*{Comparison with observations.} Although it is observationally difficult to measure the corresponding $M_\text{200}$ of a dwarf galaxy, numerical simulations \cite{onor15} suggest that a typical value is $M_\text{200} \sim 10^{10} M_\odot$. Using this as a fiducial value, we predict
	\beq
	\rho_c r_c = 41 \, M_\odot \, \text{pc}^{-2} \times \lp \frac{M_\text{200}}{10^{10} M_\odot}\rp^{0.18},
	\eeq
	where the logarithmic slope ($M^b$ with $b=0.18$) has been obtained by appropriately (log-log) linearizing the mass dependence at the fiducial value. Although the slope is calibrated at the fiducial value, it remains approximately constant over a large range of masses; even for a galaxy cluster $b = 0.22$. In particular, for the Phoenix galaxy cluster \cite{mcdonald}, $\rho_c r_c \approx 1.1 \times 10^3 \, \text{M}_\odot/\text{pc}^{-2}$. We show the full dependence on mass and redshift in Figure 2. 
	
	Remarkably, such a scaling relation has been empirically inferred by \citet{burkert15}, who reports that (in our notation) $ \rho_c r_c = 75^{+85}_{-45} \, M_\odot \, \text{pc}^{-2}$ over 18 magnitudes in blue magnitude, covering a sample that ranges from dwarf galaxies to giant galaxies. Here we have identified the central density $\rho(r=0) = \rho_c$ and the core radius $r_c$ as the length scale associated with a turnover in logarithmic slope. The reported uncertainties are not $1\sigma$ uncertainties but encompass all but 1 or 2 of the 48 data points.
	Although the median dark matter halo mass of their sample is not easily measured and therefore not reported, a reasonable value is $M = 10^{11.5} \, M_\odot$ which lies between dwarf galaxies ($M \sim 10^{10} \, M_\odot$) and giant galaxies ($M \sim 10^{13}$). Taking $M = 10^{11.5} M_\odot$ and $z=0$, we have from equation (7) that $\rho_c r_c = 78 \, M_\odot \, \text{pc}^{-2}$, which fully consistent with the value that is empirically inferred. 

The predicted scaling of $\rho_c r_c$ with mass can also be tested. 
Since the luminosity $L_{*}$ of a galaxy is proportional to the number (and thus the mass $M_{*}$ of stars, $L_{*} \propto M_{*}$. For low mass galaxies $M_{200} \lesssim 10^{12.1} M_\odot$, observations \citep{zu15} indicate $M_{200} \propto M_{*}^\beta$ with $\beta = 0.33^{+0.21}_{-0.15}$, with the slope gradually steepening to $ \propto M_{*}^{0.75}$ for very massive galaxies. 
Hence, for low mass $M_{200} \lesssim 10^{12.1} M_\odot$ (intermediate mass $M_{200} \sim 10^{12.1} M_\odot$) galaxies, we predict that $\rho_c r_c \propto M_{200}^{0.18} \propto M_{*}^{0.06} \propto L_{*}^{\gamma}$ with $\gamma = 0.06$ ($\gamma = 0.10$), 
values which are consistent with the observed slope in \cite{korm}: $\rho_c r_c \propto L_{*}^{\gamma}$, 
where $\gamma = 0.058 \pm 0.067$. Future observations, particularly of a sample of massive galaxies where the $M_{*}$ -- $L_{*}$ relation is steeper, could further constrain $\gamma$ in order to provide a more stringent test of our predictions. We also note that our model predicts virtually no redshift evolution for $z \lesssim 1$. Hence a detection of redshift dependence in $\rho_c r_c$ would falsify our model.

	It is worth mentioning that our derivation of the scaling relations allows us to compute the expected scatter in the observed relations. The two sources of scatter in our model are the scatter in $c_\text{200}$ that exists even for a fixed mass and redshift, and the scatter in $M$ if the observed population of dwarf galaxies contains galaxies of different virial masses. Note that from equation (7), $\rho_c r_c \propto c_\text{200}^2$ for large values of $c_\text{200}$, whereas the explicit mass dependence is much weaker. A typical amount of scatter (at the $1\sigma$ level) associated with $c_{200}$ is $\sim 0.1$ to $\sim 0.2$ dex \cite{dutton14}. Unless the standard deviation in the virial masses of an observed population of dwarf galaxies is greater than $\sim 1$ to $2$ orders of magnitude, the scatter in $c_{200}$ will dominate. To compare to empirical results, we will assume that the population of dwarf spheroidal galaxies observed in \cite{burkert15} does not contain such a diversity of masses. Certainly if the scatter in mass were much larger than an order of magnitude, the fact that all of their halos have $\log M_{0.3} = 7.1 \pm 0.3$ where $M_{0.3}$ is the mass enclosed in the inner 0.3 kpc would seem peculiar, since the total mass is expected to be a strong function of $M_{0.3}$. In particular, for an NFW profile gives leads to a mass dependence \cite{strig08} $M_{200} \propto \lp M_{0.3}\rp^{2.9}$, though deviations from $\rho \propto r^{-1}$ in the inner regions will change this dependency. Assuming that $c_{200}$ has a $1\sigma$ scatter of 0.15 dex, we predict using Monte Carlo methods that $\rho_c r_c = 78^{+33}_{-23} \, M_\odot \, \text{pc}^{-2}$ with a $68\%$ confidence interval. Considering that the uncertainties reported in \citet{burkert15} contain $\sim 96\%$ of the data and are thus close to $2 \sigma$ bounds, there is good agreement between the predicted and observed scatter.
	
	Our scaling laws are also in agreement with the older results of \citet{spano08} and \citet{donato09}, which find in their sample of galaxies and dwarf galaxies $\rho_c r_c = 10^{2.15 \pm 0.2} \text{M}_\odot \, \text{pc}^{-2}$ \cite{donato09}. The somewhat higher value (by a factor of $\sim 2$) of $\rho_c r_c$ compared to the results of \citet{burkert15} could be due to the fact that their sample is dominated by more massive halos. % is naturally explained by our results, since their sample is dominated by more massive halos. If the average $M_{200}$ of their sample were $\sim 10^{12} M_\text{odot}$, we would expect a $\sim 2$ factor of increase in $\rho_c r_c$.
	
	As a consistency check, we may derive additional scaling relations from our model. For example, if we assume that $r_c \ll r_s$ (an assumption that we so far have not yet made use of), the density in the region $r \lesssim r_c$ is approximately constant, and
	\beq
	\frac{M_c}{r_c^2} \approx \frac{4 \pi}{3} \rho_c r_c  \approx \text{const},
	\eeq 	
	and the core velocity dispersion $\sigma_c$ obeys $\sigma_c^2 r_c^{-1} \sim GM(r<r_c)/r_c^2 \sim G \rho_0 r_c$. Hence,
	\beq
	\sigma^2 r_c^{-1} \approx \text{const.}
	\eeq 
	%Of course, these are really just variants of the basic scaling law for $\rho_c r_c$, but they serve as a check of the consistency of our model as well as the experimental methods used to map $\rho(r)$. 
	The last two scaling laws have also been empirically inferred in \cite{burkert15} and have been shown to be consistent with available data. However, we caution that these additional results can only be derived in our formalism if $r \lesssim r_c$, which is not the general case.
	
	\paragraph*{Conclusion.} Starting from an NFW profile and a simple treatment of annihilation, we have derived a universal scaling relation $\rho_c r_c \approx 41 \, M_\odot \, \text{pc}^{-2} \times \lp M_\text{200}/{10^{10} M_\odot}\rp^{0.18}$ at $z=0$. Said differently, it has been remarked that anomalies with $\Lambda$CDM are associated with an acceleration scale $a_0 \sim 10^{-9}\, \text{cm} \, \text{s}^{-2}$ \cite{walkerloeb} where new physics becomes relevant. In units where $c=G=1$, a characteristic acceleration is equivalent to a characteristic surface density. In this work, we have derived the surface density scale (see equation 9), assuming that the new physics is dark matter self-interactions. 

	Remarkably, the derivation did not involve any additional parameters beyond what is needed in LCDM. This scaling relation holds independently of the dark matter mass and annihilation or scattering cross sections as well as the amount of adiabatic expansion $f$ experienced by the core. Our results are also relatively insensitive to slow motions of baryonic matter in the dark matter halo, provided that the effect of the baryons is to adiabatically expand or contract the core. This scaling relation is thus a robust signature of self-interaction, independent of the detailed properties of the dark matter particles. Specializing to the case of dwarf galaxies, we have shown that both the magnitude of $\rho_c \, r_c$ and the scatter are in excellent agreement with the data. We have also checked the model by deriving additional scaling relations which also agree with experiment. The predicted evolution of halo properties with redshift and mass can be tested with future observations.

\paragraph*{Acknowledgements.} This work was supported in part by NSF grant AST-1312034.

%\bibliography{cusp}{}
%
%
\end{document}